# LEARNED LOSSLESS IMAGE COMPRESSION WITH COMBINED AUTOREGRESSIVE MODELS AND ATTENTION MODULES

*Ran Wang[1]\*, Jinming Liu[1]\*, Heming Sun[2,3], Jiro Katto[1,2]*

1. Department of Computer Science and Communication Engineering, Waseda University, Tokyo, Japan
2. Waseda Research Institute for Science and Engineering, Waseda University, Tokyo, Japan
3. JST, PRESTO, 4-1-8 Honcho, Kawaguchi, Saitama, Japan

## ABSTRACT

Lossless image compression is an essential research field in image compression. Recently, learning-based image compression methods achieved impressive performance compared with traditional lossless methods, such as WebP, JPEG2000, and FLIF. However, there are still many impressive lossy compression methods that can be applied to lossless compression. Therefore, in this paper, we explore the methods widely used in lossy compression and apply them to lossless compression. Inspired by the impressive performance of the Gaussian mixture model (GMM) shown in lossy compression, we generate a lossless network architecture with GMM. Besides noticing the successful achievements of attention modules and autoregressive models, we propose to utilize attention modules and add an extra autoregressive model for raw images in our network architecture to boost the performance. Experimental results show that our approach outperforms most classical lossless compression methods and existing learning-based methods.

## 1. INTRODUCTION

Image compression is an important task in many research fields. Either lossy or lossless compression tries to capture the spatial correlations of the image to reduce the spatial redundancies in the compressed bitstream. Traditional compression standards rely on hand-crafted encoder/decoder block diagrams to reduce redundancy, such as JPEG2000 [1], WebP [2], and FLIF [3]. With the development of deep learning in recent years, learning-based image compression methods have achieved a better performance than classical compression methods.

For learning-based lossy compression, some proposed works choose to use RNN-based methods, such as [4], [5]. Some works adopt CNNs schemes, which can be interpreted as VAEs. In [7] [8], the residual block is utilized and lifted for the final performance. [9] proposed generalized divisive normalization (GDN), which obtains an impressive performance in the computer vision field. Since the end-to-end architecture of convolutional autoencoder was introduced for image compression by Ballé et al. [10], there has been plenty of novel following works, such as [11] [12]. In [11], a hierarchical autoencoder is added to a hyperprior entropy model. [12] generalizes the GSM model and adds an autoregressive context model with the hyperprior. The context model usually employs mask convolution to aggregate local contexts for efficient entropy coding. Inspired by the noticeable performance of attention modules achieved, a simplified attention module and a GMM model are utilized in [13].

As for learned-based lossless compression methods, [14] [15] obtain an impressive performance. L3C [14] that can propose a fully parallel hierarchical probabilistic model can outperform WebP, JPEG2000, and PNG. RC [15] leverages BPG to obtain a lossy reconstruction and uses the proposed RC (Residual Compressor) network to achieve lossless compression. However, there are still many effective lossy compression methods that can be utilized for lossless compression. Therefore, in this paper, we extend a learning-based lossy image compression into a lossless method.

The main contributions of this paper are listed as follows:

- We generate a Gaussian mixture model (GMM) for the latent representations and add an autoregressive component for raw images to enhance the performance.

- We apply simplified attention modules to our network architecture. With the help of attention modules, more important features have been allocated more bits that are beneficial for lifting the final performance of compression.

- We test our method on DIV2K [21], CLICP, and CLICM [22]. The result shows that our method improves up to 48% and 5.2% performance compared with PNG [6] and L3C [14], respectively.

---

\*.First two authors contributed equally.

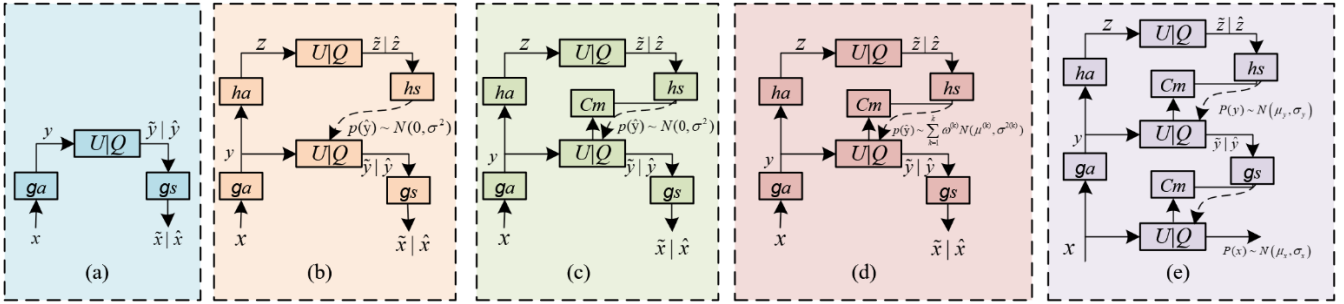

Fig.1. Operational diagrams of different learning-based methods. (a) Architecture of baseline [10]; (b) Hyperprior [11]; (c) Joint [12]; (d) Cheng [16]; (e) Proposed Model.

## 2. RELATED WORK

### 2.1 Lossless Compression Algorithms

The wildly used classical lossless compression methods include PNG [6], WebP [2], JPEG2000 [1], and FLIF [3]. FLIF is the current state-of-the-art non-learned algorithm. It relies on a well-designed entropy coding method called MANIAC. RC [15] develops a fully parallelizable non-autoregressive probabilistic learned lossless compression system that can achieve better performance compared with classical systems PNG, WebP and JPEG2000. L3C [14] is a lossless method based on the classical state-of-the-art lossy compression algorithm BPG. Except outperforms PNG, WebP, and JPEG2000, it also outperforms the non-learned state-of-the-art method FLIF on images from the Open Images data set.

### 2.2 Autoregressive Models in Image Compression

The autoregressive model is not only widely used in image compression but also in image generation and super-resolution. PixcelCNN [24] is one of the impressive works of the autoregressive model. PixcelCNN [24] estimates the joint distribution $p(x)$ of the current pixel conditions on all the previously generated pixels left and above the current pixel. Inspired by the success of autoregressive priors in probabilistic generative models, [12] introduced the context model into image compression. The context model proposed in [12] can capture spatial correlations and is able to eliminate the redundancies among the latents. It combines the context model and the hyper-network for correcting the context-based predictions.

### 2.3 Parameterized Model

According to information theory literature, in lossless compression, the input data could be considered as a stream of symbols $x_1, \ldots, x_N$, where each $x_i$ is an element from the same finite set $X$. It is important to choose a suitable distribution $p_i$ for every symbol to encode the symbol stream into a bitstream, such that we can recover the exact symbols with $p_i$ when decoding. There are several previous works with respect to parameterized models to enhance performance. [24] utilized a softmax on discrete pixel values. In [25], a discretized logistic mixture likelihood was used on the pixels, rather than a 256-way softmax, which can speed up training. Learning-based methods [14] and [15] use a discretized logistic mixture, while [12], [13], and [16] utilized discretized Gaussian mixture model to parameterize the distribution of the latent. Although which distribution can better represent the true distribution is still an open question, we follow [16] to model the latent with a Gaussian mixture model (GMM) in this paper.

## 3. METHOD

### 3.1. Formulation of Learned Compression Models

In the framework of a classical learned-based lossy image compression, the operation can be formulated as:

$$y = g_a(x; \phi)$$
$$\hat{y} = Q(y) \quad (1)$$
$$\hat{x} = g_s(\hat{y}; \theta)$$

where $x$ stands for raw images, $\hat{x}$ is reconstructed images. A latent presentation before quantization and compressed codes are denoted as $y$ and $\hat{y}$, respectively. $\phi$ and $\theta$ denote the parameters that need to be optimized for analysis and synthesis transforms. $Q$ denotes the quantization, and $U|Q$ represents quantization and entropy coding. The operation diagram is shown in Fig.1(a). In [11], a hyperprior entropy model is proposed by adding a hierarchical autoencoder, and a fully factorized density model is used for hyperprior $\hat{z}$, shown in Fig.1(b). A conditional Gaussian scale mixture model with zero-mean and scale parameters $\sigma^2$ was also introduced to the model $\hat{y}$ in [11]. Based on [11], an enhanced entropy model with an autoregressive context model is proposed in [12]. In addition, a mean and scale hyperprior is also applied in [12]. The scheme is illustrated in Fig.1(c). Following that, [13] proposed a learning-based image compression with a discretized Gaussian mixture likelihoods and attention modules. The scheme can be

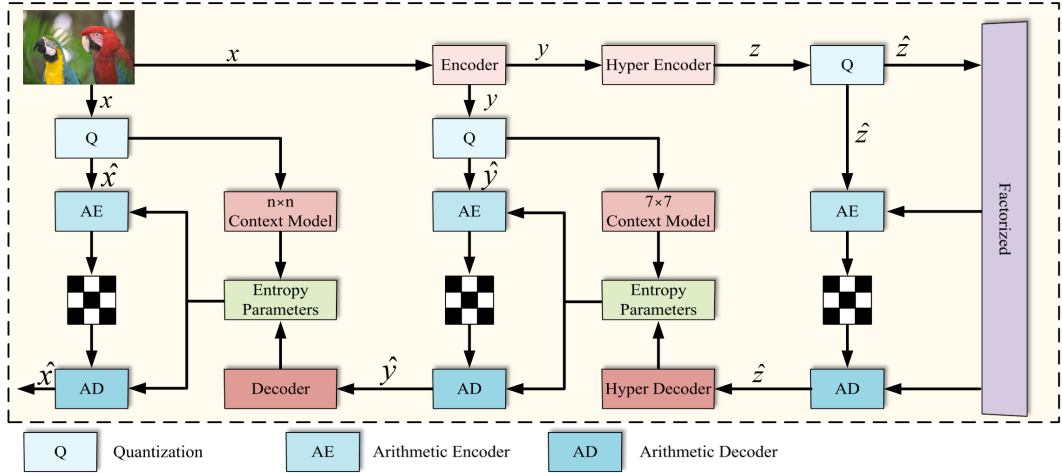

Fig.2. Network architecture.

presented in Fig.4 (d), and the entropy model is formulated as:

$$p_{\hat{y}|\hat{x}}(\hat{y}_i|\hat{z}) = \left(\sum_{k=1}^{K} w_i^{(k)} N\left(\mu_i^{(k)}, \sigma_i^{2(k)}\right) * u\left(-\frac{1}{2},\frac{1}{2}\right)\right)(\hat{y}_i) \quad (2)$$
$$= c\left(\hat{y}_i + \frac{1}{2}\right) - c\left(\hat{y}_i - \frac{1}{2}\right)$$

where $i$ represents the location in feature maps. $N$ represents a normal distribution. $u\left(-\frac{1}{2},\frac{1}{2}\right)$, $c(\cdot)$, and $k$ denote the uniform noises, the cumulative function, and the index of Gaussian mixtures, respectively. Each mixture is characterized by $w_i^{(k)}$, $\mu_i^{(k)}$, $\sigma_i^{2(k)}$, where $w_i^{(k)}$, $\mu_i^{(k)}$, and $\sigma_i^{2(k)}$ stand for weights, means and variances, respectively. $K$ is the total number of the Gaussian distributions we used. Inspired by the noticeable performance that lossy methods have achieved, we generalize a lossless model as Fig. 1(e). Our architecture is based on [16], in which the parametric form of latent distribution is improved from GSM to GMM. This paper follows the idea introduced in [16] to apply GMM distribution for $\hat{x}$.

Compared with lossy compression, lossless compression does not need to optimize the rate-distortion performance. Hence the loss function can be rewritten as:

$$L = E\left[-log_2\left(p_{\hat{x}}(\hat{x}|\mu_x, \sigma_x)\right)\right]$$
$$+ E\left[-log_2\left(p_{\hat{y}}(\hat{y}|\mu_y, \sigma_y)\right)\right] \quad (3)$$
$$+ E[-log_2(p_{\hat{z}}|\psi)]$$

where $\mu_x$, $\sigma_x$ is conditioned on $\hat{y}$. The reconstructed image $\hat{x}$ is estimated by $\mu_x$ and $\sigma_x$. $\psi$ are learnable factorized distributions.

### 3.2. Autoregressive model for lossless compression

The concept of the context model is introduced into lossy image compression by [12]. Context model is an autoregressive model over latents which utilizes mask convolution to aggregate local contexts for efficient entropy coding. In this section, we explore the idea of adding an autoregressive model to $\hat{x}$ to find a model that can fit the real distribution of images as closely as possible such that we can use fewer bits to encode images. Specifically by adding a 7 × 7 masked layer for $\hat{x}$ in our proposed approach. The expression of $\hat{x}$ is shown in Eq.5,

$$P_{\hat{x}|\hat{y}}(\hat{x}|\hat{y}) \sim \sum_{k=1}^{K} w^k N(\mu^{(k)}, \sigma^{2(k)}) \quad (4)$$

where $k$ denotes the index of mixtures. Each mixture is characterized by a Gaussian distribution with $w, \mu, \sigma$ 3 parameters, which represent for weights, means and variances.

## 4. EXPERIMENTS

### 4.1. Training Details

We compare our method with different classical and learning-based lossless works. The proposed network architecture is shown in Fig.2. We add a masked layer to the latent $\hat{y}$ and reconstructed image $\hat{x}$, respectively. We also insert attention modules into the encoder-decoder for our network. For training, $K$ is set as 3 and $\lambda$ sets as 0.6 for the first 20 epochs for faster convergence. After 20 epochs, we choose to set $\lambda$ as 0. This L2-norm shares the same idea with mean square error (MSE) loss in lossy image compression.

We chose about 40000 patches from the ImageNet [19] and cropped the size to 256×256 before randomly fed into the network. These images are not ideal for the lossless compression task, but we are not aware of a suitable lossless training data set. The number of filters N was set to 192 in our network. The model was optimized using Adam [20] with a batch size of 8, and the learning rate was set to 1× 10⁻⁴ at the beginning of training. After 100 epochs, the learning rate

Table 1. Compression performance on different datasets.

| [bpsp] | CLICP | | CLICM | | DIV2K | |
|---|---|---|---|---|---|---|
| Ours | 2.839 | | 2.632 | | 2.942 | |
| Cheng [16] | 3.355 | +18.2% | 3.231 | +22.8% | 3.476 | +18.2% |
| RC [15] | 2.933 | +3.3% | 2.538 | -3.6% | 3.079 | +4.7% |
| L3C [14] | 2.944 | +3.7% | 2.639 | +0.3% | 3.094 | +5.2% |
| JPEG2000 | 3.000 | +5.7% | 2.721 | +3.4% | 3.127 | +6.3% |
| WebP | 3.006 | +5.9% | 2.774 | +5.4% | 3.176 | +8.0% |
| FLIF | 2.784 | -1.9% | 2.492 | -5.3% | 2.911 | -1.1% |
| PNG | 3.997 | +40.8% | 3.896 | +48.0% | 4.235 | +43.9% |

dropped to $1\times10^{-5}$. Our model was trained 400 epochs to obtain stable performance.

Table 2. Performance of different masked layers.

| | CLICP | CLICM | DIV2K |
|---|---|---|---|
| Cheng [16] + Attention | 3.282 | 3.134 | 3.392 |
| Cheng [16] + Attention + 5×5 masked | 2.942 | 2.759 | 3.064 |
| **Cheng [16] + Attention + 7×7 masked** | **2.839** | **2.632** | **2.942** |

### 4.2. Evaluation

For comparison, we evaluate our model on three datasets, CLICP, CLICM, and DIV2K. The rate is measured by bits per sub-pixel (bpsp). It can be observed from Table 1 that our method outperforms the classical methods for all datasets, such as PNG [6], JPEG2000 [1] and WebP [2], while marginally worse than FLIF [3]. It is noticeable that FLIF involves many highly specialized hand-crafted techniques. Furthermore, compared with widely-used PNG, our method achieves over 40% improvement for all three test datasets. As for JPEG2000 and WebP, our method achieves about 5% improvement. As for the learning-based methods, we lift about 20% and 3% performance compared with [16] and L3C [14], respectively. Our method achieves an improvement of about 3.3% and 4.7% in CLICP and DIV2K compared with RC [15] but a little worse for the CLICM dataset.

### 4.3. Ablation Study

**Autoregressive model** From Table 2, it can be observed that the model with an extra autoregressive model for $\hat{x}$ lifts the performance of compression. The reason is that the autoregressive model for the reconstructed image $\hat{x}$ reduces uncertainties in the meantime, avoiding introducing more information that needs to be compressed. We also explore different configurations for the autoregressive model. The performance comparison is shown in Table 2. It can be observed that network architecture with the masked layer for the reconstructed image $\hat{x}$ achieves better performance, and the network with the 7 × 7 masked layer can outperform the 5 × 5 masked layer slightly. This is in line with our perception that a larger masked layer may achieve better performance.

**Attention modules** Attention mechanisms have the potential to improve the coding efficiency by capturing the global correlations. In this paper, we employ simplified attention modules proposed in [13]. To solely verify the effectiveness of the attention modules utilized in our network, we evaluated the compression effect of the network of [16] with and without attention modules on CLICP, CLICM, and DIV2K, respectively. The original network of [16] is without attention modules. It is worth noting that the only difference between the two models is whether attention modules are added. As shown in Table 3, the network with attention modules achieve better performance.

Table 3. Performance of with/without attention modules.

| | CLICP | CLICM | DIV2K |
|---|---|---|---|
| Cheng [16] | 3.355 | 3.231 | 3.476 |
| **Cheng [16] + Attention** | **3.282** | **3.134** | **3.392** |

### 5. CONCLUSION

In this paper, we improve the performance of the learning-based lossless compression method by adding attention modules and combining an autoregressive model for the reconstructed image $\hat{x}$. Experiments demonstrate that our proposed method outperforms most classical lossless image compression methods, such as PNG, JPEG2000, and WebP. For learning-based methods, we outperform L3C and [16] for all test datasets and are slightly worse than RC for the CLICM dataset. To further improve our work, we could investigate different forms of context models to capture global-scope spatial correlations and cross-channel relationships among the latents. In addition, we could explore different types of attention modules to build a more powerful network.